\documentclass[conference]{IEEEtran}
\IEEEoverridecommandlockouts
\usepackage{cite}
\usepackage{amsmath,amssymb,amsfonts}
\usepackage{algorithmic}
\usepackage{graphicx}
\usepackage{textcomp}
\usepackage{xcolor}
\usepackage{bm}
\allowdisplaybreaks[4]
\newtheorem{theorem}{Theorem}

\newtheorem{lemma}{Lemma}

\newtheorem{remark}{Remark}

\def\BibTeX{{\rm B\kern-.05em{\sc i\kern-.025em b}\kern-.08em
    T\kern-.1667em\lower.7ex\hbox{E}\kern-.125emX}}

\makeatletter
\newcommand{\linebreakand}{%
	\end{@IEEEauthorhalign}
	\hfill\mbox{}\par
	\mbox{}\hfill\begin{@IEEEauthorhalign}
	}
\makeatother

\begin{document}

\title{DoF Upper Bound on Three-User MISO Broadcast Channel with  Two Antennas and Mixed CSIT
}

\author{\IEEEauthorblockN{1\textsuperscript{st} Shuo Zheng}
	\IEEEauthorblockA{\textit{Department of EEE}\\ 
		\textit{Southern University of }\\ \textit{Science and Technology}\\
		Shenzhen, China \\
		zhengs2021@mail.sustech.edu.cn}
\and
\IEEEauthorblockN{2\textsuperscript{nd} Tong Zhang}
\IEEEauthorblockA{\textit{College of Information}\\ \textit{Science and Technology} \\
	\textit{Jinan University}\\
	Guangzhou, China \\
	zhangt77@jnu.edu.cn}
\and
\IEEEauthorblockN{3\textsuperscript{rd} Shuai Wang}
\IEEEauthorblockA{\textit{Shenzhen Institute of}\\ \textit{ Advanced Technology} \\
	\textit{Chinese Academy of Sciences}\\
	Shenzhen, China \\
	s.wang@siat.ac.cn}
\and
\IEEEauthorblockN{4\textsuperscript{th} Gaojie Chen}
\IEEEauthorblockA{\textit{5GIC \& 6GIC, Institute} \\ \textit{ for Communication Systems} \\
	\textit{University of Surrey}\\
	GU2 7XH Guildford, U.K. \\
	gaojie.chen@surrey.ac.uk}
\linebreakand
\IEEEauthorblockN{5\textsuperscript{th} Guoliang Li}
\IEEEauthorblockA{\textit{Department of EEE}\\ 
	\textit{Southern University of }\\ \textit{Science and Technology}\\
	Shenzhen, China \\
	ligl2020@mail.sustech.edu.cn}
\and
\IEEEauthorblockN{6\textsuperscript{th} Shancheng Zhao}
\IEEEauthorblockA{\textit{College of Information}\\ \textit{Science and Technology} \\
	\textit{Jinan University}\\ 
	Guangzhou, China \\
	shanchengzhao@jnu.edu.cn}
\and
\IEEEauthorblockN{7\textsuperscript{th} Rui Wang}
\IEEEauthorblockA{\textit{Department of EEE} \\
\textit{Southern University of} \\ \textit{ Science and Technology}\\
Shenzhen, China \\
wang.r@sustech.edu.cn}
}

\maketitle

\begin{abstract}
With delayed and imperfect current channel state information at the transmitter (CSIT), namely mixed CSIT, the sum degrees-of-freedom (sum-DoF) in the two-user multiple-input multiple-output (MIMO) broadcast channel  and the $K$-user multiple-input single-output (MISO) broadcast channel with not-less-than-$K$ transmit antennas have been obtained. However, the case of the three-user broadcast channel with two transmit antennas and mixed CSIT is still unexplored. In this paper, we investigate the sum-DoF upper bound of three-user MISO broadcast channel with two transmit antennas and mixed CSIT. By exploiting genie-aided signaling and extremal inequalities, we derive the sum-DoF upper bound as $(1-\alpha)3/2 + 9\alpha/4$, which is  at most $12.5\%$ larger than the expected sum-DoF, given by $(1-\alpha)3/2 + 2\alpha$. This indicates that the gap may mitigate by better bounding the imperfect current CSIT counterpart.
\end{abstract}

\begin{IEEEkeywords}
DoF, mixed CSIT, upper bound, three-user MISO broadcast channel  
\end{IEEEkeywords}

\section{Introduction}
It has been recognized that channel state information at the transmitter (CSIT) is very important to communication systems. Unfortunately, perfect CSIT is always desirable, but oftentimes intractable. There can be two practical scenarios having delayed CSIT, namely CSIT reflect only past state information of channel. One case is high-mobility communications, where the channel state information (CSI) feedback is lagging behind the CSI varying \cite{Ai}. The other case is satellite communications, where the large communication latency makes the CSIT feedback delayed \cite{Satellite}. To understanding the fundamental performance limits for networks with delayed CSIT, characterizing the degrees-of-freedom (DoF), which is a first-order approximation of capacity when signal-to-noise ratio (SNR) is high, was therefore considered  \cite{Maddah-Ali,Vaze,25,26,28}.

As a seminal work, the DoF region of the $K$-user multiple-input single-output (MISO) broadcast channel with delayed CSIT was characterized in \cite{Maddah-Ali}. Subsequently, the DoF region of the two-user multiple-input multiple-output (MIMO) broadcast channel with delayed CSIT was derived in \cite{Vaze}. For the DoF region of the three-user MIMO broadcast channel with delayed CSIT, related works can be found in \cite{25,26,28}.

Shifting from stringent requirement on delayed CSIT, there can be a mix of delayed and imperfect current CSIT, so called mixed CSIT. This is because the wireless channels are temporally correlated, and the transmitter can estimate partial current CSIT from delayed CSIT due to this temporal correlation. With mixed CSIT, the DoF region of two-user MISO broadcast channel was characterized in \cite{ShengYang} and \cite{Gou}. Then,   the DoF region of the two-user broadcast channel and interference channel with mixed CSIT was derived in \cite{Yi}. For the two-user MIMO interference channel with mixed CSIT, a unified precoding strategy was proposed in \cite{Rezaee}. The DoF region of two-user Z MIMO interference channel with mixed CSIT was derived in \cite{Mohanty}. For multi-hop networks, the DoF region of two-hop MISO broadcast channel with mixed CSIT was obtained by \cite{Wangzhao}.  Recently, for the $K$-user MISO broadcast channel with mixed CSIT when the number of transmit antennas is not less than $K$, the sum-DoF were characterized in \cite{Kerret}. It was found that this sum-DoF in $K$-user broadcast channel is indeed a convex combination of the sum-DoF of delayed CSIT and the sum-DoF of imperfect current CSIT. However, although many intriguing results have been reported, back to the three-user MISO broadcast channel  with two transmit antennas and mixed CSIT, there is still unclear for the sum-DoF.

In this paper, we study the sum-DoF upper bound for the three-user MISO broadcast channel with mixed CSIT, when number of transmit antennas is two. This sum-DoF upper bound is derived via the standard converse techniques. More specifically, we use the same genie-aided signaling and extremal inequalities as in \cite{ShengYang, Yi,Kerret,de2016optimal}. We expect the sum-DoF is a convex combination of the sum-DoF of delayed CSIT and the sum-DoF of imperfect current CSIT. However, this sum-DoF upper bound has a gap from the expected sum-DoF, rather than the tightness when the number of transmit antennas is  large.  Fortunately, this gap is  at most $12.5\%$ larger than the expected sum-DoF. This result encourages us to design a better sum-DoF upper bound, which may improve by better bounding the imperfect current CSIT counterpart.

\section{System Model}

\begin{figure}[t]
	\centering
	\includegraphics[width=.95\linewidth]{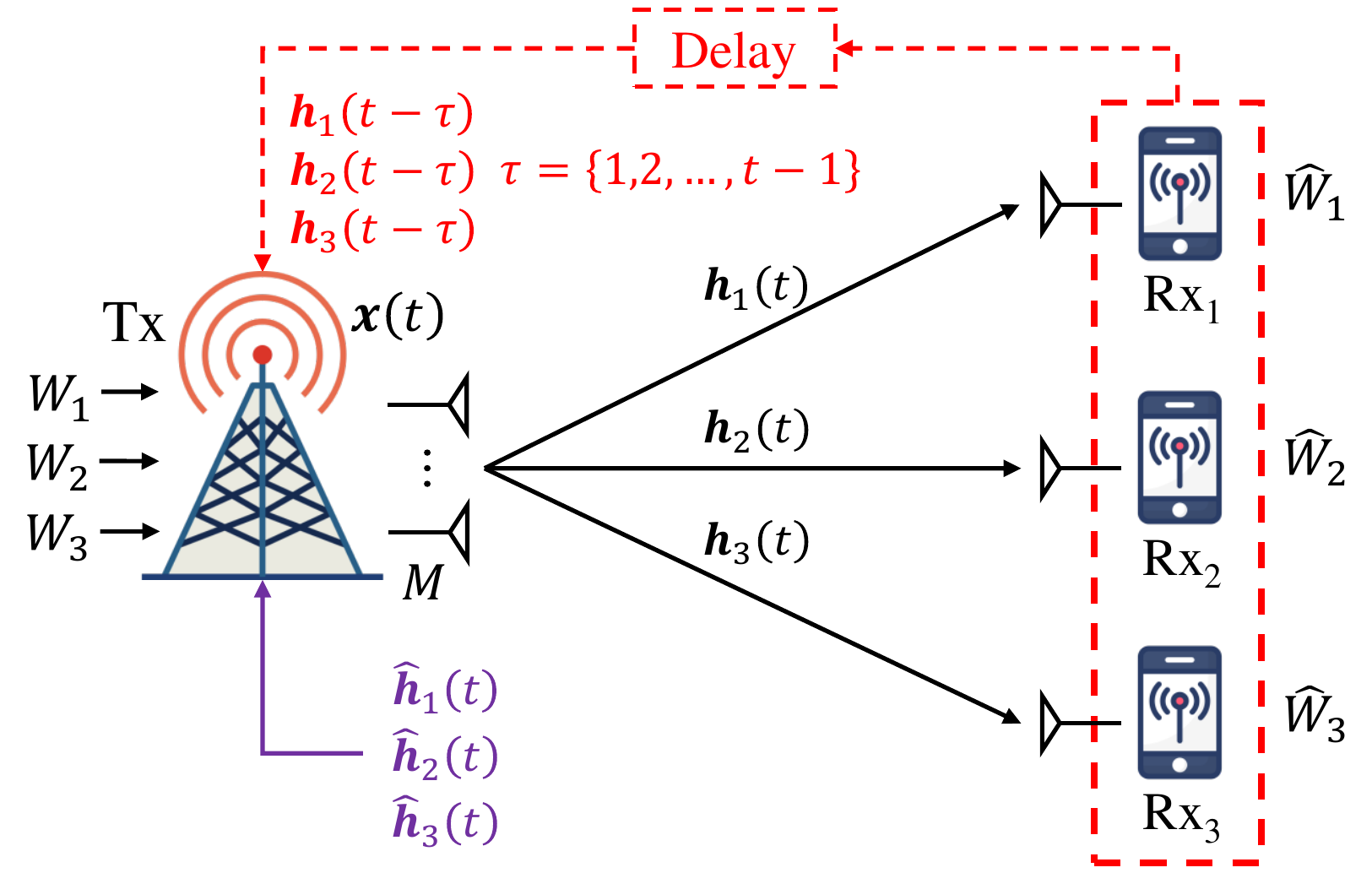}
	\caption{System model of three-user MISO broadcast channel with mixed CSIT, where the transmitter has $M$ antennas and each receiver has one antenna.} 
\end{figure}
\begin{figure}[t]
	\centering
	\includegraphics[width=.4\linewidth]{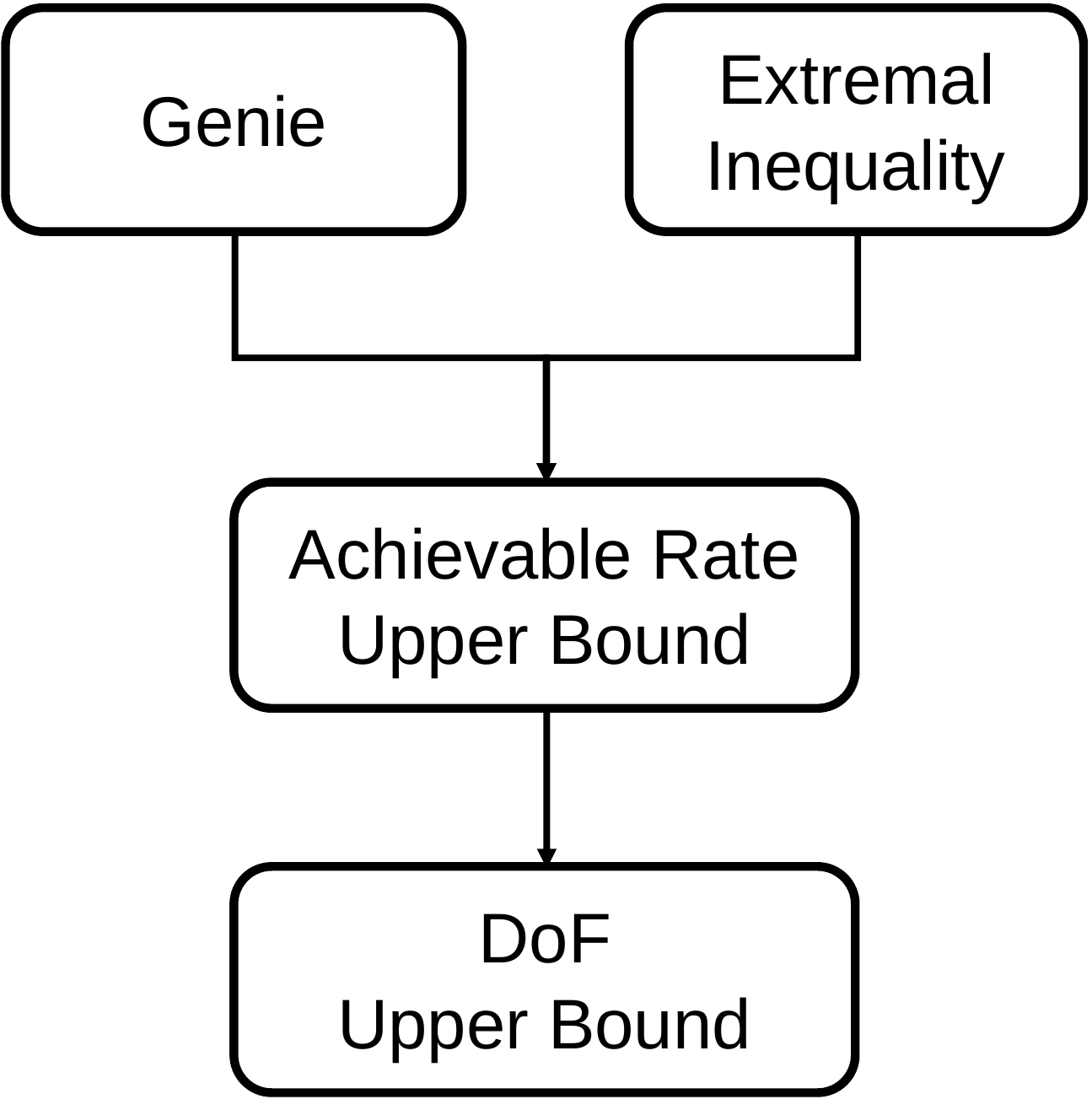}
	\caption{Flow chart of the upper bound proof for $M=2$.}
\end{figure}

As shown in Fig. 1, we consider a three-user MISO broadcast channel, where the transmitter equipped with $M$ antennas is denoted by $\mathrm{Tx}$ and three single-antenna receivers are denoted by $\mathrm{Rx}_1$, $\mathrm{Rx}_2$ and $\mathrm{Rx}_3$, respectively. 
The received signal at the $
\mathrm{Rx}_i,\,i=1,2,3,$ is expressed as
\begin{align}
	y_i(t)=\bm{h}_i^H(t)\bm{x}(t)+z_i(t),
\end{align}
where $t$ represents the time slot, $\bm{h}_i^H(t)\in \mathbb{C}^{1\times M}$ represents the channel vector for $\mathrm{Rx}_i$, $\bm{x}(t)\in \mathbb{C}^{M\times 1}$ is the transmitted signal subject to power constraint $\mathbb{E}\left(\Vert \bm{x}(t) \Vert^2\right)\leq P$,  $z_i(t)\sim \mathcal{CN}(0,1)$ is the additive white Gaussian noise (AWGN), which is independent of channel vectors and the transmitted signal. 
For convenience, we further define $\mathcal{H}(t)\triangleq [\bm{h}_1(t),\bm{h}_2(t),\bm{h}_3(t)]^H \in \mathbb{C}^{3 \times M}$ and $\mathcal{H}^n\triangleq \{\mathcal{H}(t)\}_{t=1}^n$.	
At the $t$-th time slot, $\mathrm{Tx}$ is assumed to know the historical delayed CSIT $\mathcal{H}^{t-1}$. Therefore, $\mathrm{Tx}$ can produce an imperfect estimate of the current CSI $\widehat{\mathcal{H}}(t)$. The channel estimate is modeled as
\begin{align}
	\bm{h}_i(t)=\widehat{\bm{h}}_i(t)+\widetilde{\bm{h}}_i(t),
\end{align}
where the channel estimate $\widehat{\bm{h}}_i(t)\sim \mathcal{N}_{\mathbb{C}}(0,(1-\sigma^2)\mathbf{I})$ and the channel estimate error $\widetilde{\bm{h}}_i(t) \sim \mathcal{N}_{\mathbb{C}}(0,\sigma^2\mathbf{I})$ are mutually independent, where $\mathbf{I}$ denotes the identity matrix.
$\mathcal{H}(t)$ is independent of $(\mathcal{H}^{t-1},\widehat{\mathcal{H}}^{t-1})$ when conditioned on $\widehat{\mathcal{H}}(t)$. Besides, $\mathrm{Rx}$s are assumed to know $\mathcal{H}^t$ and $\widehat{\mathcal{H}}^t$ after the transmission at time slot $t$. According to \cite{Yi}, CSIT quality is defined as
\begin{align}
	\alpha = -\lim\limits_{P\rightarrow \infty}\frac{\log \sigma^2}{\log P}.
\end{align}
Under this definition, $\mathbb{E}[\vert \bm{h}^H_i\bm{w} \vert^2]\sim P^{-\alpha}$ given that $\bm{w}$ is a normalized zero-forcing precoding vector (i.e., $\widehat{\bm{h}}_i^H\bm{w}=0$), where $\alpha = 0$ means that only delayed CSIT is available, while $\alpha \rightarrow \infty$ corresponds to the case with perfect current CSIT. Since $\alpha > 1$ is sufficient to achieve the maximum DoF, $\alpha \in [0,1]$ is analyzed in this article from the DoF perspective.

Denote $W_i$ the message for $\mathrm{Rx}_i$ and $R_i=\frac{\log \vert W_i \vert}{n}$ the rate, where $\vert W_i\vert$ is the cardinality of the corresponding message set. The rate tuple $(R_1(P),R_2(P),R_3(P))$ is said to be achievable if there exists a coding scheme such that the average decoding error probability $P_e^{(n)}$ for all messages approaches zero as the channel use $n$ goes to infinity. The sum-capacity $C_{\Sigma}(P)$ is defined to be the supremum of all achievable sum-rates. Therefore, the sum-DoF is defined as $
d_1 + d_2 + d_3 = \lim_{P\rightarrow \infty}\frac{C_{\Sigma}}{\log P}.
$

\begin{figure*}[htbp]
	\begin{align}
		&\frac{n(R_1-\mathcal{O}(1))}{2}+\frac{n(R_2-\mathcal{O}(1))}{2}+n(R_3-\mathcal{O}(1))\notag\\
		& \overset{(a)}{\leq} \sum_{t=1}^n \left\{ \frac{1}{2}h(\bm{Y}_{[1:3]}(t)| \mathcal{U}_1 (t),\mathcal{H}(t)) -\frac{1}{2}h(\bm{Y}_{[1:3]}(t)|W_1, \mathcal{U}_1 (t),\mathcal{H}(t))+\frac{1}{2}h(\bm{Y}_{[2:3]}(t)|\mathcal{U}_2 (t),\mathcal{H}(t))\right.\nonumber\\
		&\hspace{1.8cm} \left.-\frac{1}{2}h(\bm{Y}_{[2:3]}(t)|W_2, \mathcal{U}_2 (t),\mathcal{H}(t))
		+h(y_3(t)|\mathcal{U}_3 (t),\mathcal{H}(t)) -h(y_3(t)|W_3, \mathcal{U}_3 (t),\mathcal{H}(t))  \right\} \nonumber \\
		& =\sum_{t=1}^n\left\{ \frac{1}{2}h(\bm{Y}_{[1:3]}(t)| \mathcal{U}_1 (t),\mathcal{H}(t)) -\frac{1}{2}h(\bm{Y}_{[2:3]}(t)|W_2, \mathcal{U}_2 (t),\mathcal{H}(t))+\frac{1}{2}h(\bm{Y}_{[2:3]}(t)|\mathcal{U}_2 (t),\mathcal{H}(t))\right.\nonumber\\
		&\hspace{1.8cm} \left.
		-\frac{}{}\! h(y_3(t)|W_3, \mathcal{U}_3 (t),\mathcal{H}(t))  \right \}+h(y_3(t)|\mathcal{U}_3 (t),\mathcal{H}(t))-\frac{1}{2}h(\bm{Y}_{[1:3]}(t)|W_1, \mathcal{U}_1 (t),\mathcal{H}(t)) \nonumber \\
		& \overset{(b)}{\leq} \sum_{t=1}^n\left\{ \frac{1}{2}h(\bm{Y}_{[1:3]}(t)| \mathcal{U}_1 (t),\mathcal{H}(t))-\frac{1}{2}h(\bm{Y}_{[2:3]}(t)|W_2, \mathcal{U}_2 (t),\mathcal{H}(t),\bm{Y}_1^{t-1}) +\frac{1}{2}h(\bm{Y}_{[2:3]}(t)|\mathcal{U}_2 (t),\mathcal{H}(t))\right.\nonumber\\
		&\hspace{1.8cm} \left. \!\!\frac{}{}-h(y_3(t)|W_3, \mathcal{U}_3 (t),\mathcal{H}(t),\bm{Y}_2^{t-1})\right\}+n\log P +n \mathcal{O}(1) \nonumber\\
		& = \sum_{t=1}^n\left\{ \frac{1}{2}h(\bm{Y}_{[1:3]}(t)| \mathcal{U}_1 (t),\mathcal{H}(t))-\frac{1}{2}h(\bm{Y}_{[2:3]}(t)|\mathcal{U}_1 (t),\mathcal{H}(t)) +\frac{1}{2}h(\bm{Y}_{[2:3]}(t)|\mathcal{U}_2 (t),\mathcal{H}(t))- h(y_3(t)|\mathcal{U}_2 (t),\mathcal{H}(t))\right\}\nonumber\\
		&\hspace{1.8cm} +n\log P +n \mathcal{O}(1)\label{wRate} \tag{8}
	\end{align}\hrule
	\begin{align}
		& \frac{1}{2} h\left(\bm{Y}_{[1:3]}(t)| \mathcal{U}_1(t), \mathcal{H}(t)\right)-\frac{1}{2} h\left(\bm{Y}_{[2:3]}(t)| \mathcal{U}_1(t), \mathcal{H}(t)\right) \nonumber\\
		&\hspace{1cm} \leq \mathbb{E}_{\widehat{\mathcal{H}}(t)} \max _{\substack{\bm{D} \succeq 0 \\
				\operatorname{tr}(\bm{D}) \leq P}} \mathbb{E}_{\mathcal{H}(t) | \widehat{\mathcal{H}}(t)}\left(\frac{1}{2} \log \operatorname{det}\left(\mathbf{I}+\mathcal{H}_{[1:3]}(t) \bm{D}(t) \mathcal{H}_{[1:3]}^{H}(t)\right) -\frac{1}{2} \log \operatorname{det}\left(\mathbf{I}+\mathcal{H}_{[2: 3]}(t) \bm{D}(t) \mathcal{H}_{[2: 3]}^{H}(t)\right)\right)\label{ratebound1}\tag{9}\\
		& \frac{1}{2} h\left(\bm{Y}_{[2:3]}(t) | \mathcal{U}_2(t), \mathcal{H}(t)\right)-h\left(y_3(t) | \mathcal{U}_2(t), \mathcal{H}(t)\right) \nonumber\\
		&\hspace{1cm} \leq \mathbb{E}_{\widehat{\mathcal{H}}(t)} \max _{\substack{\bm{D} \succeq 0 \\
				\operatorname{tr}(\bm{D}) \leq P}} \mathbb{E}_{\mathcal{H}(t) | \widehat{\mathcal{H}}(t)}\left(\frac{1}{2} \log \operatorname{det}\left(\mathbf{I}+\mathcal{H}_{[2:3]}(t) \bm{D}(t) \mathcal{H}_{[2:3]}^{H}(t)\right)-\log \left(1+\bm{h}^H_3(t) \bm{D}(t) \bm{h}_3(t)\right)\right)\label{ratebound2}\tag{10}
	\end{align}
	\hrule
\end{figure*}

\section{Main Results and Discussion}\label{AA}
\begin{theorem}
	For the three-user MISO broadcast channel with two transmit antennas and mixed CSIT, defined in Section-II, the sum-DoF upper bound is given by
\begin{equation}
	d_1+d_2+d_3 \le (1-\alpha)\frac{3}{2} + \frac{9}{4}\alpha.
\end{equation}
\end{theorem}

\subsection{Proof of Theorem 1}
For three transmit antennas, the sum-DoF upper bound is matched with the sum-DoF lower bound, which was derived in \cite{de2016optimal}. For one transmit antenna, the sum-DoF is trivial. Therefore, it remains to investigate the sum-DoF upper bound with two transmit antennas.

In this case,  our proof of the upper bound is established on standard technique given in \cite{ShengYang, Yi,Kerret,de2016optimal}, including
\begin{enumerate}
	\item Genie-based bounding technique to upper bound the achievable rates \cite{ShengYang,Yi}.
	\item Extremal inequality to upper bound the entropy difference terms \cite{32}. 
\end{enumerate}
Specifically, as shown in Fig. 2, we first apply the same genie-aided signaling and then utilize the same extremal inequality as in \cite{ShengYang, Yi,Kerret,de2016optimal} to bound the achievable rates.

Below, please find our derivation of the upper bound when antenna configuration is $M=2$.

A genie provides $\mathrm{Rx}_i$ with $\mathrm{Rx}_j$'s message $W_j$ as well as the received signals $y_j(m),\forall m\le t$, where $j> i$. For convenience, we denote notations $\bm{Y}_{[i:j]}(t)\triangleq\{y_k(t)\}_{k=i}^{j}$, $\bm{Y}_{[i:j]}^n\triangleq \{\bm{Y}_{[i:j]}(t)\}_{t=1}^n$, $\mathcal{H}_{[i:j]}(t)\triangleq [\bm{h}_i(t),\bm{h}_{i+1}(t),...,\bm{h}_{j}(t)]^H$ and  $\bm{z}_{[i:j]}(t)\triangleq [z_i(t),z_{i+1}(t),...,z_j(t)]^H$, where $j\ge i$.

\begin{figure*}[tbp]
	\begin{align}
		& \frac{1}{2} h\left(\bm{Y}_{[1:3]}(t) | \mathcal{U}_1(t), \mathcal{H}(t)\right)-\frac{1}{2} h\left(\bm{Y}_{[2:3]}(t) | \mathcal{U}_1(t), \mathcal{H}(t)\right)\nonumber \\
		&\hspace{0.3cm} \leq \max _{p\left(\mathcal{U}_1(t)\right), p\left(\bm{x}(t) | \mathcal{U}_1(t)\right)}\left(\frac{h\left(\bm{Y}_{[1:3]}(t) | \mathcal{U}_1(t), \mathcal{H}(t)\right)}{2}-\frac{h\left(\bm{Y}_{[2:3]}(t) | \mathcal{U}_1(t), \mathcal{H}(t)\right)}{2}\right) \nonumber\\
		&\hspace{0.3cm} \overset{(a)}{\leq} \max_{p\left(\mathcal{U}_1(t)\right)} \mathbb{E}_{\mathcal{U}_1(t)} \max _{p\left(\bm{x}(t) | \mathcal{U}_1(t)\right)}\left(\frac{h\left(\bm{Y}_{[1:3]}(t)| \mathcal{U}_1(t), \mathcal{H}(t)\right)}{2}-\frac{h\left(\bm{Y}_{[2:3]}(t) |\mathcal{U}_1(t), \mathcal{H}(t)\right)}{2}\right) \nonumber\\
		&\hspace{0.3cm} =\max _{p\left(\mathcal{U}_1(t)\right)} \mathbb{E}_{\mathcal{U}_1(t)} \max _{p\left(\bm{x}(t) | \mathcal{U}_1(t)\right)} \mathbb{E}_{\mathcal{H}(t) | \mathcal{U}_1(t)}\left(\frac{h\left(\bm{Y}_{[1:3]}(t) | \mathcal{U}_1(t), \mathcal{H}(t)\right)}{2}-\frac{h\left(\bm{Y}_{[2:3]}(t) | \mathcal{U}_1(t), \mathcal{H}(t)\right)}{2}\right) \nonumber\\
		&\hspace{0.3cm} \overset{(b)}{=}\max _{p\left(\mathcal{U}_1(t)\right)} \mathbb{E}_{\mathcal{U}_1(t)} \max _{p\left(\bm{x}(t) | \mathcal{U}_1(t)\right)} \mathbb{E}_{\mathcal{H}(t) | \hat{\mathcal{H}}(t)}\left(\frac{h\left(\mathcal{H}_{[1:3]}(t) \bm{x}(t)+\bm{z}_{[1:3]}(t) | \mathcal{U}_1(t)\right)}{2}-\frac{h\left(\mathcal{H}_{[2:3]}(t) \bm{x}(t)+\bm{z}_{[2:3]}(t) | \mathcal{U}_1(t)\right)}{2}\right) \nonumber\\
		&\hspace{0.3cm} \overset{(c)}{=}\max _{p\left(\mathcal{U}_1(t)\right)} \mathbb{E}_{\mathcal{U}_1(t)} \max _{\substack{\bm{D} \succeq 0 \\
				\operatorname{tr}(\bm{D}) \leq P}} \max _{\substack{p\left(\bm{x}(t) | \mathcal{U}_1(t)\right) \\
				\operatorname{cov}\left(\bm{x}(t) | \mathcal{U}_1(t)\right) \preceq \bm{D}}} \mathbb{E}_{\mathcal{H}(t) | \hat{\mathcal{H}}(t)}\left(\frac{h\left(\mathcal{H}_{[1:3]}(t) \bm{x}(t)+\bm{z}_{[1:3]}(t) | \mathcal{U}_1(t)\right)}{2}\right.\nonumber \\
		&\hspace{9cm} \left.-\frac{h\left(\mathcal{H}_{[2:3]}(t) \bm{x}(t)+\bm{z}_{[2:3]}(t) | \mathcal{U}_1(t)\right)}{2}\right)\nonumber \\
		& \hspace{0.3cm}\overset{(d)}{\leq} \max _{p\left(\mathcal{U}_1(t)\right)} \mathbb{E}_{\mathcal{U}_1(t)} \max _{\substack{\bm{D} \succeq 0 \\
				\operatorname{tr}(\bm{D}) \leq P}} \max _{\bm{K}(t) \preceq \bm{D}} \mathbb{E}_{\mathcal{H}(t) | \hat{\mathcal{H}}(t)}\left(\frac{1}{2} \log \operatorname{det}\left(\mathbf{I}+\mathcal{H}_{[1:3]}(t) \bm{K}(t) \mathcal{H}_{[1:3]}^{H}(t)\right)\right. \nonumber\\
		&\hspace{8cm} \left.-\frac{1}{2} \log \operatorname{det}\left(\mathbf{I}+\mathcal{H}_{[2:3]}(t) \bm{K}(t) \mathcal{H}_{[2:3]}^{H}(t)\right)\right) \nonumber\\
		&\hspace{0.3cm} =\max _{p\left(\mathcal{U}_1(t)\right)} \mathbb{E}_{\mathcal{U}_1(t)} \max _{\substack{\bm{D} \succeq 0 \\
				\operatorname{tr}(\bm{D}) \leq P}} \mathbb{E}_{\mathcal{H}(t) | \hat{\mathcal{H}}(t)}\left(\frac{1}{2} \log \operatorname{det}\left(\mathbf{I}+\mathcal{H}_{[1:3]}(t) \bm{K}^*(t) \mathcal{H}_{[1:3]}^{H}(t)\right)-\frac{1}{2} \log \operatorname{det}\left(\mathbf{I}+\mathcal{H}_{[2:3]}(t) \bm{K}^*(t) \mathcal{H}_{[2:3]}^{H}(t)\right)\right)\nonumber \\
		&\hspace{0.3cm} \overset{(e)}{\leq} \mathbb{E}_{\hat{\mathcal{H}}(t)} \max _{\substack{\bm{D} \succeq 0 \\
				\operatorname{tr}(\bm{D}) \leq P}} \mathbb{E}_{\mathcal{H}(t) | \hat{\mathcal{H}}(t)}\left(\frac{1}{2} \log \operatorname{det}\left(\mathbf{I}+\mathcal{H}_{[1:3]}(t) \bm{D}(t) \mathcal{H}_{[1:3]}^{H}(t)\right)-\frac{1}{2} \log \operatorname{det}\left(\mathbf{I}+\mathcal{H}_{[2:3]}(t) \bm{D}(t) \mathcal{H}_{[2:3]}^{H}(t)\right)\right) \label{end}\tag{11}
	\end{align}
	\hrule
\end{figure*}

Employing Fano's inequality, we can upper bound the achievable rate of $\mathrm{Rx}_1$ as
\begin{align}
	&n(R_1-\mathcal{O}(1))\nonumber\\
	&\overset{(a)}{\leq} I(W_1;W_{[2:3]},\bm{Y}_{[1:3]}^n|\mathcal{H}^n,\widehat{\mathcal{H}}^n)\nonumber\\
	&\overset{(b)}{=}I(W_1;\bm{Y}_{[1:3]}^n|W_{[2:3]},\mathcal{H}^n,\widehat{\mathcal{H}}^n)\nonumber\\
	&\overset{(c)}{=}\sum\limits_{t=1}^n I(W_1;\bm{Y}_{[1:3]}(t)|W_{[2:3]},\bm{Y}_{[1:3]}^{t-1},\mathcal{H}^n,\widehat{\mathcal{H}}^n)\nonumber\\
	&\overset{(d)}{=}\sum_{t=1}^n \left(h(\bm{Y}_{[1:3]}(t)|W_{[2:3]},\bm{Y}_{[1:3]}^{t-1}, \mathcal{H}^t,\widehat{\mathcal{H}}^t)\right.\nonumber\\
	&\quad \quad \quad\left.-h(\bm{Y}_{[1:3]}(t)|W_{[1:3]},\bm{Y}_{[1:3]}^{t-1},\mathcal{H}^t,\widehat{\mathcal{H}}^t)\right)\nonumber\\
	&=\sum_{t=1}^n \left(h(\bm{Y}_{[1:3]}(t)| \mathcal{U}_1(t),\mathcal{H}(t))\right.\nonumber\\
	&\quad\quad \quad \left.-h(\bm{Y}_{[1:3]}(t)|W_1,\mathcal{U}_1 (t),\mathcal{H}(t))\right)\label{result1}
\end{align}
where $\mathcal{U}_1 (t)\triangleq \left\{ W_{[2:3]},\bm{Y}_{[1:3]}^{t-1},\mathcal{H}^{t-1},\widehat{\mathcal{H}}^t\right\}$; (a) holds by using Fano's inequality; (b) holds due to the independence of messages; (c) holds by using the chain rule of mutual information; (d) holds due to the definition of mutual information.

The achievable rate of $\mathrm{Rx}_2$ is bounded as
\begin{align}
	& n(R_2-\mathcal{O}(1)) \nonumber\\
	&\overset{(a)}{\leq} I(W_2;W_3,\bm{Y}_{[2:3]}^n|\mathcal{H}^n,\widehat{\mathcal{H}}^n)\nonumber\\
	&\overset{(b)}{=}I(W_2;\bm{Y}_{[2:3]}^n|W_3,\mathcal{H}^n,\widehat{\mathcal{H}}^n)\nonumber\\
	&\overset{(c)}{=}\sum\limits_{t=1}^n I(W_2;\bm{Y}_{[2:3]}(t)|W_3,\bm{Y}_{[2:3]}^{t-1},\mathcal{H}^n,\widehat{\mathcal{H}}^n)\nonumber\\
	&\overset{(d)}{=}\sum_{t=1}^n \left(h(\bm{Y}_{[2:3]}(t)|W_3,\bm{Y}_{[2:3]}^{t-1}, \mathcal{H}^t,\widehat{\mathcal{H}}^t)\right.\nonumber\\
	&\quad \  \quad\left.-h(\bm{Y}_{[2:3]}(t)|W_{[2:3]},\bm{Y}_{[2:3]}^{t-1},\mathcal{H}^t,\widehat{\mathcal{H}}^t)\right)\nonumber\\
	&=\sum_{t=1}^n \left(h(\bm{Y}_{[2:3]}(t)| \mathcal{U}_2(t),\mathcal{H}(t))\right.\nonumber\\
	&\quad\quad \quad \left.-h(\bm{Y}_{[2:3]}(t)|W_2,\mathcal{U}_2 (t),\mathcal{H}(t))\right)\label{result2}
\end{align}
where $\mathcal{U}_2 (t)\triangleq \left\{ W_3,\bm{Y}_{[2:3]}^{t-1},\mathcal{H}^{t-1},\widehat{\mathcal{H}}^t\right\}$; (a) holds by using Fano's inequality;
(b) holds due to independence of messages; (c) holds by using the chain rule of mutual information; (d) holds due to the definition of mutual information.

The achievable rate of $\mathrm{Rx}_3$ is bounded as
\begin{align}
	&n(R_3-\mathcal{O}(1))  \nonumber \\
	&\overset{(a)}{\leq} I(W_3;\bm{Y}_3^n|\mathcal{H}^n,\widehat{\mathcal{H}}^n) \nonumber \\
	&\overset{(c)}{=}\sum\limits_{t=1}^n I(W_3;y_3(t)|\bm{Y}_3^{t-1},\mathcal{H}^n,\widehat{\mathcal{H}}^n) \nonumber\\
	&\overset{(d)}{=}\sum_{t=1}^n \left(h(y_3(t)|\bm{Y}_3^{t-1}, \mathcal{H}^t,\widehat{\mathcal{H}}^t)\right.\nonumber\\
	&\quad \quad \quad\left.-h(y_3(t)|W_3,\bm{Y}_3^{t-1},\mathcal{H}^t,\widehat{\mathcal{H}}^t)\right) \nonumber\\
	&=\sum_{t=1}^n \left(h(y_3(t)| \mathcal{U}_3(t),\mathcal{H}(t))\right.\nonumber\\
	&\quad\quad \quad \left.-h(y_3(t)|W_3,\mathcal{U}_3 (t),\mathcal{H}(t))\right),\label{result3}
\end{align}
where $\mathcal{U}_3 (t)\triangleq \left\{ \bm{Y}_3^{t-1},\mathcal{H}^{t-1},\widehat{\mathcal{H}}^t\right\}$; (a) holds by using Fano's inequality;
(b) holds due to the chain rule of mutual information;
(c) holds due to the definition of mutual information. 

In the following, we upper bound the weighted sum-rate given by \eqref{wRate} on the top of page 3,
 where $\mathcal{O}(1)$ denotes a constant which does not scale as power $P$; (a) From \eqref{result1}, \eqref{result2} and \eqref{result3};
(b) Differential entropy decreases due to conditioning.

\begin{lemma}
	For each entropy difference term in \eqref{wRate}, we can establish upper bounds in \eqref{ratebound1} and \eqref{ratebound2}.
\end{lemma}
\begin{IEEEproof}
To avoid repetition, we only give the proof of the first entropy difference term provided in \eqref{end} shown on the top of the last page,
where $\bm{K}$ and $\bm{K}^*$ denote the covariance matrix of $\bm{x}(t)$ and the optimal $\bm{K}$, respectively. The reason of each step is given as follows:
\begin{enumerate}
	\item[(a)] The maximization is moved inside the expectation and this is not less than the original value.
	\item[(b)] $\mathcal{H}(t)$ is independent of $(\mathcal{H}^{t-1},\hat{\mathcal{H}}^{t-1})$ when conditioned on $\hat{\mathcal{H}}(t)$.
	\item[(c)] The maximization is divided into two parts (i.e., trace constraint and covariance matrix constraint).
	\item[(d)] $\bm{x}(t)$ with Gaussian distribution is optimal for maximization, which is derived via extremal inequality \cite{32}.
	\item[(e)] $\bm{K}^*$ always satisfies $\bm{K}^*\succeq 0, \operatorname{tr}(\bm{K}^*)\leq P$ and inner expectation only relates to $\hat{\mathcal{H}}(t)$, where $\operatorname{tr}(\cdot)$ denotes the trace.
\end{enumerate}
\end{IEEEproof}

According to \eqref{ratebound1}, \eqref{ratebound2} and \cite[Lemma 3]{Yi}, we have the following upper bound of the weighted rate. 
\begin{align}
	&\frac{R_1-\mathcal{O}(1)}{2}+\frac{R_2-\mathcal{O}(1)}{2}+R_3-\mathcal{O}(1)  \nonumber \\
	& \leq\frac{1}{2}\alpha \log P + \log P+\mathcal{O}(1). \nonumber
\end{align}
Therefore, we obtain
\begin{gather}
	\frac{d_1}{2}+\frac{d_2}{2}+d_3\leq 1+\frac{\alpha}{2}.\tag{12}
\end{gather}
Due to the symmetry, we have
\begin{align}
	&\frac{d_1}{2}+d_2+\frac{d_3}{2}\leq 1 + \frac{\alpha}{2},\tag{13}\\
	&d_1+\frac{d_2}{2}+\frac{d_3}{2}\leq 1 + \frac{\alpha}{2}.\tag{14}
\end{align}
By adding (10)-(12), we can obtain the desired result  
\begin{gather}
	d_1+d_2+d_3\leq \frac{3}{2}+\frac{3\alpha}{4}=\frac{3}{2}(1-\alpha) + \frac{9}{4}\alpha.\tag{15}
\end{gather}
 
\begin{figure}[t]
	\centering
	\includegraphics[width=.8\linewidth]{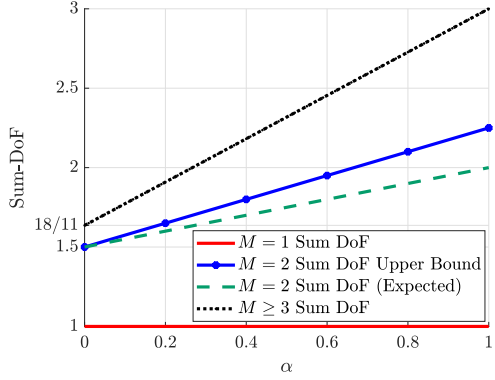} 
	\caption{Sum-DoF upper bound.}
\end{figure}
\begin{figure}[t]
	\centering
	\includegraphics[width=.8\linewidth]{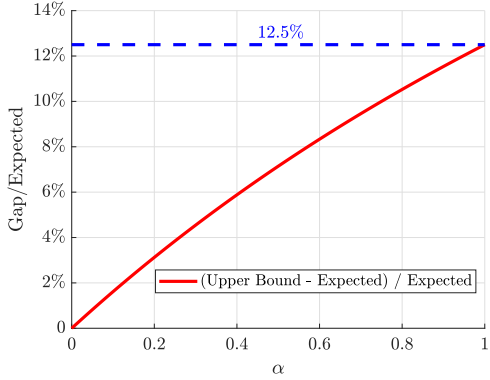}
	\caption{Gap between the upper bound and the expected.} 
\end{figure}
\subsection{Discussion}
	For $M=2$, the sum-DoF of delayed CSIT is $3/2$ and the sum-DoF of perfect current CSIT is $2$. Therefore, we expect the sum-DoF as $(1-\alpha)3/2 + 2\alpha$. However, this value is not toughed by our upper bound, as illustrated in Fig. 3. In addition, Fig. 3 shows that the gain between $M \ge 3$ Case and $M=2$ Case is increasing when $\alpha$ approaches $1$. We further depict the gap in Fig. 4, where it shows the gap increases with $\alpha$ and is at most $12.5\%$ larger than the expected sum-DoF. By comparing  our bound with expected one, we infer that the imperfect current CSIT part in bounding may be improved. 
	
	For the three-user MISO broadcast channel with mixed CSIT, the sum-DoF upper bound is now complete, given by
	\begin{equation}
		d_1+d_2+d_3 \le 
		\begin{cases}
			(1-\alpha)18/11 + 3\alpha, & M\ge 3, \\
			(1-\alpha)3/2 + 9\alpha/4, & M=2, \\
			1, & M=1.
		\end{cases} \nonumber 
	\end{equation}

	\bibliographystyle{IEEEtran}
	\bibliography{ICCCS2023}

\begin{thebibliography}{10}
\providecommand{\url}[1]{#1}
\csname url@samestyle\endcsname
\providecommand{\newblock}{\relax}
\providecommand{\bibinfo}[2]{#2}
\providecommand{\BIBentrySTDinterwordspacing}{\spaceskip=0pt\relax}
\providecommand{\BIBentryALTinterwordstretchfactor}{4}
\providecommand{\BIBentryALTinterwordspacing}{\spaceskip=\fontdimen2\font plus
\BIBentryALTinterwordstretchfactor\fontdimen3\font minus
  \fontdimen4\font\relax}
\providecommand{\BIBforeignlanguage}[2]{{%
\expandafter\ifx\csname l@#1\endcsname\relax
\typeout{** WARNING: IEEEtran.bst: No hyphenation pattern has been}%
\typeout{** loaded for the language `#1'. Using the pattern for}%
\typeout{** the default language instead.}%
\else
\language=\csname l@#1\endcsname
\fi
#2}}
\providecommand{\BIBdecl}{\relax}
\BIBdecl

\bibitem{Ai}
J.~Li, Y.~Niu, H.~Wu, B.~Ai, S.~Chen, Z.~Feng, Z.~Zhong, and N.~Wang,
  ``Mobility support for millimeter wave communications: Opportunities and
  challenges,'' \emph{{IEEE} Commun. Surveys Tuts.}, vol.~24, no.~3, pp.
  1816--1842, 2022.

\bibitem{Satellite}
O.~Kodheli, E.~Lagunas, N.~Maturo, S.~K. Sharma, B.~Shankar, J.~F.~M. Montoya,
  J.~C.~M. Duncan, D.~Spano, S.~Chatzinotas, S.~Kisseleff, J.~Querol, L.~Lei,
  T.~X. Vu, and G.~Goussetis, ``Satellite communications in the new space era:
  A survey and future challenges,'' \emph{{IEEE} Commun. Surveys Tuts.},
  vol.~23, no.~1, pp. 70--109, 2021.

\bibitem{Maddah-Ali}
M.~A. Maddah-Ali and D.~Tse, ``Completely stale transmitter channel state
  information is still very useful,'' \emph{IEEE Trans. Inf. Theory}, vol.~58,
  no.~7, pp. 4418--4431, July 2012.

\bibitem{Vaze}
C.~S. Vaze and M.~K. Varanasi, ``The degrees of freedom region of the two-user
  {MIMO} broadcast channel with delayed {CSIT},'' in \emph{Proc. IEEE Int.
  Symp. Inf. Theory (ISIT)}, July 2011, pp. 199--203.

\bibitem{25}
M.~J. Abdoli, A.~Ghasemi, and A.~K. Khandani, ``On the degrees of freedom of
  three-user {MIMO} broadcast channel with delayed {CSIT},'' in \emph{Proc.
  IEEE Int. Symp. Inf. Theory (ISIT)}, St. Petersburg, Russia, 2011, pp.
  209--213.

\bibitem{26}
T.~Zhang, X.~W. Wu, Y.~F. Xu, Y.~Ge, and P.~C. Ching, ``Three-user {MIMO}
  broadcast channel with delayed {CSIT}: {A} higher achievable {DoF},'' in
  \emph{Proc. IEEE Int. Conf. Acoust., Speech, Signal Process. (ICASSP)}, 2018,
  pp. 3709--3713.

\bibitem{28}
T.~{Zhang} and R.~{Wang}, ``Achievable {DoF} regions of three-user {MIMO}
  broadcast channel with delayed {CSIT},'' \emph{IEEE Trans. Commun.}, vol.~69,
  no.~4, pp. 2240--2253, 2021.

\bibitem{ShengYang}
S.~Yang, M.~Kobayashi, D.~Gesbert, and X.~Yi, ``Degrees of freedom of time
  correlated {MISO} broadcast channel with delayed {CSIT},'' \emph{IEEE Trans.
  Inf. Theory}, vol.~59, no.~1, pp. 315--328, 2013.

\bibitem{Gou}
T.~Gou and S.~A. Jafar, ``Optimal use of current and outdated channel state
  information: Degrees of freedom of the {MISO} {BC} with mixed {CSIT},''
  \emph{IEEE Commun. Lett.}, vol.~16, no.~7, pp. 1084--1087, 2012.

\bibitem{Yi}
X.~Yi, S.~Yang, D.~Gesbert, and M.~Kobayashi, ``The degrees of freedom region
  of temporally correlated {MIMO} networks with delayed {CSIT},'' \emph{IEEE
  Trans. Inf. Theory}, vol.~60, no.~1, pp. 494--514, 2014.

\bibitem{Rezaee}
M.~Rezaee and P.~J. Schreier, ``A degrees-of-freedom-achieving scheme for the
  temporally correlated {MIMO} interference channel with delayed {CSIT},''
  \emph{IEEE Trans. Wireless Commun.}, vol.~17, no.~8, pp. 5397--5408, 2018.

\bibitem{Mohanty}
K.~Mohanty and M.~K. Varanasi, ``Degrees of freedom region of the {MIMO}
  {Z}-interference channel with mixed {CSIT},'' \emph{IEEE Commun. Lett.},
  vol.~20, no.~12, pp. 2422--2425, 2016.

\bibitem{Wangzhao}
Z.~Wang, M.~Xiao, C.~Wang, and M.~Skoglund, ``Degrees of freedom of two-hop
  {MISO} broadcast networks with mixed {CSIT},'' \emph{IEEE Trans. Wireless
  Commun.}, vol.~13, no.~12, pp. 6982--6995, 2014.

\bibitem{Kerret}
P.~de~Kerret, D.~Gesbert, J.~Zhang, and P.~Elia, ``Optimal {DoF} of the
  {K}-user broadcast channel with delayed and imperfect current {CSIT},''
  \emph{IEEE Trans. Inf. Theory}, vol.~66, no.~11, pp. 7056--7066, 2020.

\bibitem{de2016optimal}
------, ``Optimal {DoF} of the {K}-user broadcast channel with delayed and
  imperfect current {CSIT},'' \emph{arXiv preprint arXiv:1604.01653}, 2016.

\bibitem{32}
T.~Liu and P.~Viswanath, ``An extremal inequality motivated by multiterminal
  information-theoretic problems,'' \emph{IEEE Trans. Inf Theory}, vol.~53,
  no.~5, pp. 1839--1851, 2007.

\end{thebibliography}

\end{document}